# NMR of $^3$He in contact with PrF$_3$ nanoparticles at low temperatures

**M. S. Tagirov[a], E. M. Alakshin[a], R.R. Gazizulin[a], A.V. Egorov[a],**
**A. V. Klochkov[a], S. L. Korableva[a], V.V. Kuzmin[a],**
**A.S. Nizamutdinov[a], K. Kono[b], A. Nakao[b], and A.T. Gubaidullin[c]**

*[a]Faculty of Physics, Kazan (Volga region) Federal University,*
*Kremlevskaya, 18, 420008, Kazan, Russia*
*[b]RIKEN, Wako, Saitama, 351-0198, Japan*
*[c]A.E. Arbuzov Institute of Organic and Physical Chemistry of Kazan*
*Scientific Center of Russian Academy of Sciences , Ac. Arbuzov street,*
*420088, Kazan, Russia*

*Two nanosized PrF$_3$ samples were synthesized using two different procedures. The X-ray and HRTEM experiments showed high crystallinity of synthesized sample. Comparison of enhanced $^{141}$Pr NMR spectra of microsized (45 µm) and nanosized PrF$_3$ powder is presented. Experimental data on spin kinetics of $^3$He in contact with PrF$_3$ nanoparticles at T = 1.5 K are reported.*

## 1. INTRODUCTION

The $^3$He and its magnetic properties at low temperatures are studied widely since 1970s and a lot of interesting effects was discovered, like superfluidity [1] and spin superfluidity [2]. The properties of $^3$He in porous media is the matter of interest, for example an aerogel suppress superfluid transition of $^3$He, because the length scale of an open geometry of aerogel is the same order of magnitude as the superfluid coherence length [3]. A lot of NMR experiments studied magnetic properties of adsorbed $^3$He films and surface effects in bulk $^3$He (see, for example, review sec. [4]). The magnetic properties and spin kinetics of $^3$He at temperatures above Fermi temperature of liquid $^3$He are also studied extensively. For instance, the influence of a porous media on NMR characteristics of liquid $^3$He can be used as a tool for study porous substrates [5]. Another interesting effect was discovered in 1980s, which is a magnetic coupling between liquid $^3$He and nuclear spin systems of solid state substrates [6, 7, 8]. The resonance magnetic coupling between $^3$He and van Vleck paramagnet TmEs was discovered [9].

**M. S. Tagirov[a], E. M. Alakshin[a], R.R. Gazizulin[a], A.V. Egorov[a], A. V. Klochkov[a], S. L. Korableva[a], V.V.Kuzmin[a], A.S.Nizamutdinov[a], K. Kono[b], A. Nakao[b], and A.T. Gubaidullin[c]**

The "PrF$_3$–liquid $^3$He" system is of interest because of the possibility of using the magnetic coupling between the nuclei of the two spin systems for the dynamic nuclear polarization of liquid $^3$He [10]. Van Vleck paramagnets are known to have high anisotropy of the effective nuclear magnetogyric ratio [11]. As a result, a direct interaction between magnetic moments of equal magnitude at the liquid $^3$He–solid state substrate interface becomes possible.

The resonance magnetic coupling between liquid $^3$He nuclei and the $^{141}$Pr nuclei of microsized (45 μm) Van Vleck paramagnet PrF$_3$ powder has been discovered [12,13]. Using nanosized PrF$_3$ powder would create a highly-coupled $^3$He - $^{141}$Pr spin system and could show new aspects of effects discovered earlier.

## 2. THE SAMPLES

The nanosized PrF$_3$ samples were synthesized by using the methods described in [14]. In a typical synthesis, 2.48 g of praseodymium oxide was dissolved in 160 ml of a 10% nitric acid solution to form a transparent solution, then 1.9 g of NaF (F:Pr=3:1,) was added into the above solution under violent stirring. A light green colloidal precipitate of PrF$_3$ appeared immediately. The pH of the suspension was adjusted by ammonia to about 4.0–5.0. Deionized water was filled into the suspension to make the volume up to 300 ml. After stirring for about 20 min, the suspension was finally transferred into a 500 ml round flask and placed in a microwave oven (650 W, 2.45 GHz). The suspension was heated by microwave irradiation for 20 min at 70% of the maximum power under refluxing. The resulting product was collected by centrifugation and washed several times using deionized water and ethanol. Thus, two samples were obtained: sample A (without microwave radiation), sample B (with microwave radiation).

The crystal structure of the samples has been characterized by X-ray diffraction (XRD) (Fig. 1). 1-st line – sample A, 2-nd line - sample A (seven weeks aging at STP), 3-d line - sample B. All of the diffraction peaks can be readily indexed from the standard powder diffraction data of the hexagonal phase PrF$_3$. Fig. 1 also confirms the hexagonal phase of the of PrF$_3$ particles crystal structure. As shown in Fig. 1, the narrow and sharp peaks indicate high crystallinity of the samples.

High-resolution transmission electron microscopy (HRTEM) images were obtained by using JEM – 2100 F/SP with resolution – 0,14 nm  using an accelerating voltage of 200 kV (Fig.2 and Fig. 3).

**NMR of $^3$He in contact with PrF$_3$ nanoparticles at low temperatures**

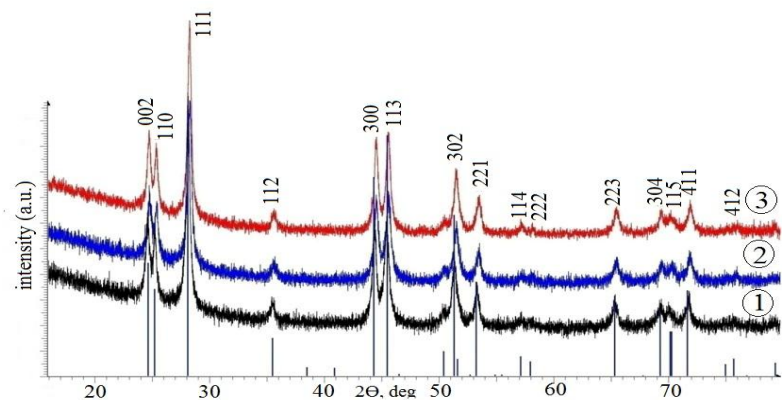

Fig. 1. XRD of PrF$_3$ samples. 1-st line – sample A, 2-nd line - sample A seven weeks old (checking for aging), 3-d line - sample B.

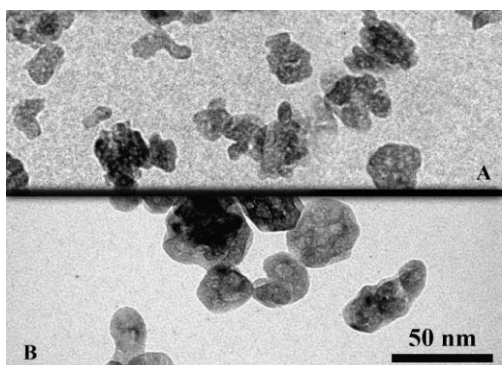

Fig. 2. HRTEM image of PrF$_3$ nanoparticles

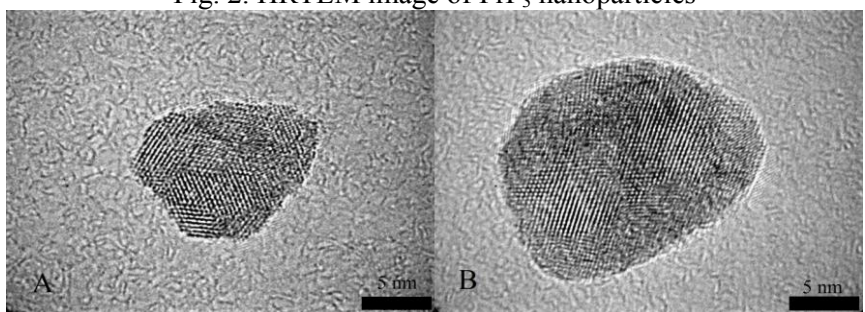

Fig. 3. HRTEM image of an individual PrF$_3$ nanoparticle

As shown in Fig. 2, nanoparticles of sample B are larger than those of sample A and have more regular spherical shape. Fig. 3 shows that transition from sample A to sample B leads to transition from polycrystalline to single

**M. S. Tagirov[a], E. M. Alakshin[a], R.R. Gazizulin[a], A.V. Egorov[a], A. V. Klochkov[a], S. L. Korableva[a], V.V.Kuzmin[a], A.S.Nizamutdinov[a], K. Kono[b], A. Nakao[b], and A.T. Gubaidullin[c]**

crystal structure. From HRTEM images we can obtain a size distribution of $PrF_3$ nanoparticles for samples A and B (Fig. 4). Particle sizes are: $20 \pm 15$ nm sample A, $32 \pm 10$ nm sample B.

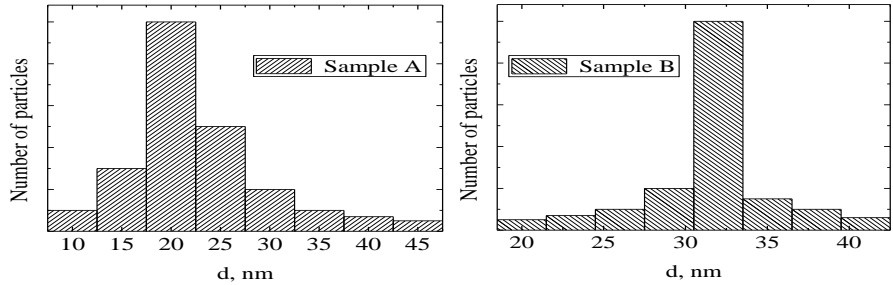

Fig. 4. Distribution of $PrF_3$ nanoparticles size for samples A and B

### 3. EXPERIMENT AND RESULTS

In all presented experiments the synthesized $PrF_3$ sample B was used, because it has more uniform particles and narrow size distribution. The sample (m = 345 mg) was placed in a glass tube which was sealed leak tight to the $^3He$ gas handling system. On the outer surface of the glass tube an NMR coil was mounted. A homebuilt pulse NMR spectrometer (frequency range of 3–50 MHz) was used. The pulse NMR spectrometer is equipped with a resistive electrical magnet that has a magnetic field strength up to 1 T. All experiments were done at the temperature T=1.5 K, which was achieved by helium bath pumping. The longitudinal magnetization relaxation time $T_1$ of $^3He$ was measured by the saturation recovery method using a spin-echo signal.

In the experiments with $^3He$ three different systems were studied: $PrF_3$ nanoparticles completely covered by $^3He$ adsorbed layer, $PrF_3$ sample filled by gas phase $^3He$ and $PrF_3$ sample filled by liquid $^3He$. The amount of $^3He$, necessary for complete coverage of $PrF_3$ nanoparticles was adjusted as in [15] and was equal to 10 $cm^3$ STP for our sample. Gaseous $^3He$ was condensed into the sample cell at temperature T=1.5 K in small portions on the order of 0.5 $cm^3$ STP. After the condensation of each portion, the pressure in the sample cell was checked and, if it turned out to be less than $10^{-2}$ mbar, the next portion was condensed. The entire sample surface was assumed to be coated by the adsorbed layer of $^3He$ atoms when the equilibrium pressure exceeded $10^{-1}$ mbar.

The NMR spectra of $^{141}Pr$ ($I= 5/2$) observed in a $PrF_3$ single crystal are well described by the nuclear spin Hamiltonian [16]:

**NMR of ³He in contact with PrF₃ nanoparticles at low temperatures**

$$H = -\hbar \sum_{i=x,y,z} \gamma_i H_i I_i + D\left[I_z^2 - \frac{1}{3}I(I+1)\right] + E\left(I_x^2 - I_y^2\right)$$ (1)

where $\gamma_x/2\pi = 3.32(2)$ kHz/Oe, $\gamma_y/2\pi = 3.24(2)$ kHz/Oe, $\gamma_z/2\pi = 10.03(5)$ kHz/Oe, $|D/h| = 4.31(1)$ MHz, and $|E/h| = 0.30(1)$ MHz.

The resonance NMR spectra of the powdered PrF₃ sample were measured at frequencies of 6.63 and 19.5 MHz (Fig. 5) by method described in [12] (filled points - PrF₃ nanoparticles, open points- microsized particles [12], the NMR cell was filled by liquid ⁴He for a good thermal contact with the external helium bath). The simulated NMR spectra of ¹⁴¹Pr for the PrF₃ powders presented in Fig.6.

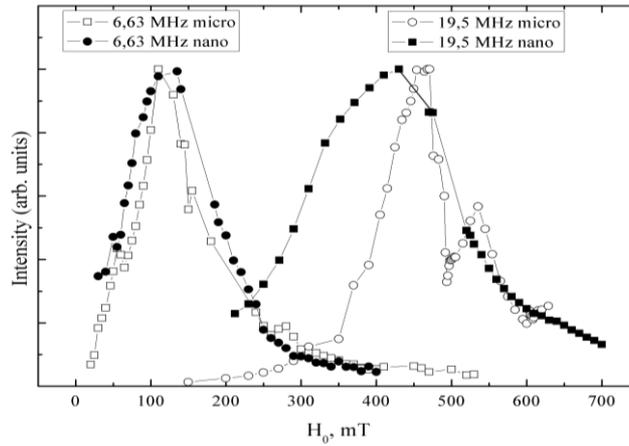

Fig. 5. Enhanced nuclear magnetic resonance (ENMR) spectra of ¹⁴¹Pr in PrF₃ powder obtained at a frequency 6.63 and 19.5 MHz (filled points - PrF₃ nanoparticles (sample B), open points- microsized particles[12])

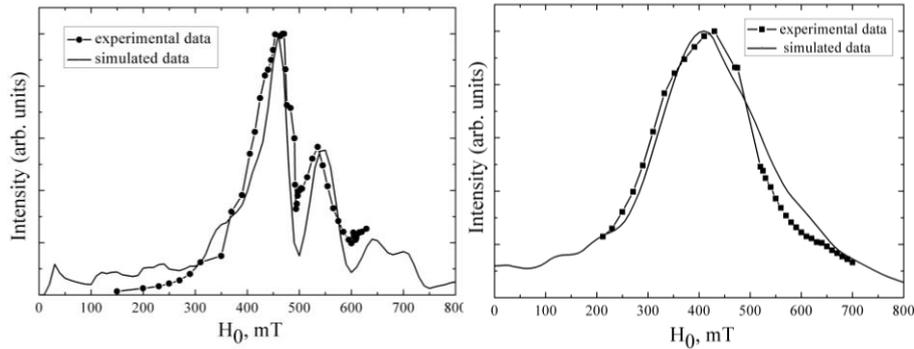

Fig. 6. Simulated nuclear magnetic resonance spectra of ¹⁴¹Pr in PrF₃ powder recorded at the frequency 19.5 MHz. Microsized – left fig., nanoparticles - right fig.

**M. S. Tagirov[a], E. M. Alakshin[a], R.R. Gazizulin[a], A.V. Egorov[a], A. V. Klochkov[a], S. L. Korableva[a], V.V.Kuzmin[a], A.S.Nizamutdinov[a], K. Kono[b], A. Nakao[b], and A.T. Gubaidullin[c]**

Fig. 6 shows that at a frequency of 19.5 MHz the nanosized sample spectrum is well described by the simulated spectrum with a line width 6 MHz, and microsized - 1 MHz. [141]Pr energy levels in $PrF_3$ in the absence of a magnetic field are 9.02 and 17.08 MHz. Fig. 7 shows that with the transition from single crystal to the microsized powder, the levels become a zone of a width of 1 MHz, and the transition to nanosized sample widens the zone up to 6 MHz.

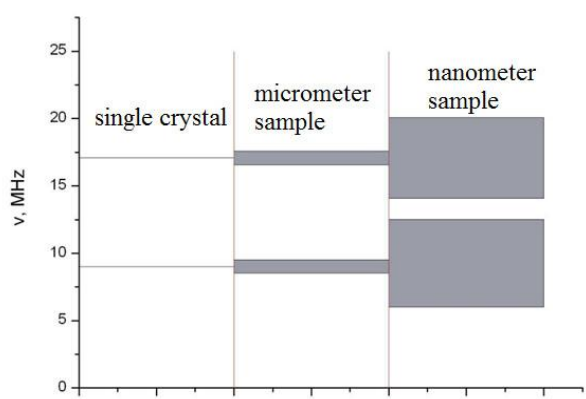

Fig. 7. Schematic of [141]Pr energy levels in $PrF_3$ in the absence of applied magnetic field vs. size of sample particles.

The spin kinetics of $^3$He in contact with $PrF_3$ nanoparticles has been studied. Experimental data are presented in Fig.8 and Fig.9. In all experimental cases ($PrF_3$ nanoparticles completely covered by $^3$He adsorbed layer, $PrF_3$ sample filled by gas phase $^3$He and $PrF_3$ sample filled by liquid $^3$He) only one common $^3$He NMR signal was detected. This situation is similar to $^3$He NMR in an aerogel samples at low [15,17] and ultralow temperatures [18] and could be result of fast exchange of $^3$He atoms between adsorbed layer and free (gas or liquid) phase.

The magnetic field dependences of the longitudinal magnetization relaxation rate of $^3$He nuclei measured by pulse NMR technique in the system $PrF_3$ - $^3$He are presented in Fig. 8. As can be seen the nuclear magnetic relaxation of $^3$He is anomalously fast. In normal bulk liquid $^3$He the longitudinal relaxation time is the order of hundreds of seconds [19].

Nanosized powder has a huge crystal surface which means that relaxation mechanisms due to the adsorbed film [20,21,15] should work. Due to the fact that $PrF_3$ is a van Vleck paramagnet, its particles create a highly inhomogeneous magnetic field between them. Thus, the longitudinal

**NMR of $^3$He in contact with PrF$_3$ nanoparticles at low temperatures**

relaxation mechanism of $^3$He in inhomogeneous field should work [22] and this model assumes the relaxation rate should be proportional to the inhomogeneity of magnetic field between powder particles. According to our measurements of $T_2^*$ versus magnetic field (Fig.9), the inhomogeneity of magnetic field is proportional to external magnetic field $H_0$. Taking into account both described mechanisms all experimental data can be fitted by following equation (parameters A and B presented in Table 1.):

$$1/T_1 = A/H_0 + BH_0 \qquad (2)$$

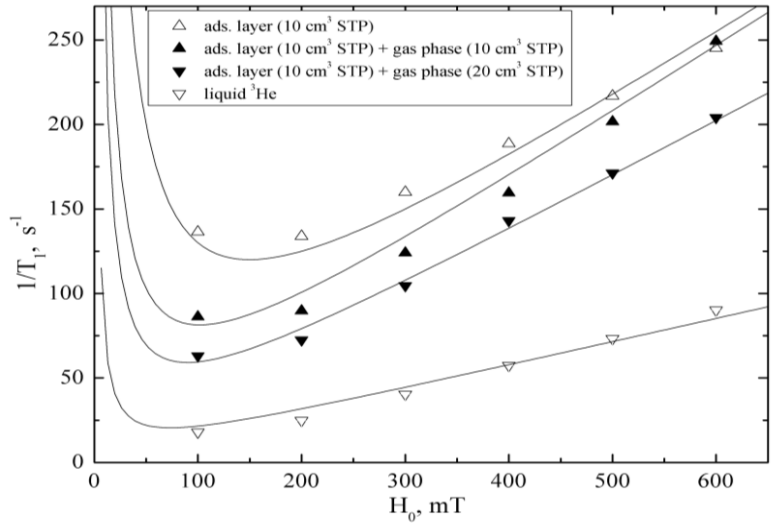

Fig. 8. The magnetic field dependences of the longitudinal magnetization relaxation rate of $^3$He nuclei in the system PrF$_3$ - $^3$He.

| The system under investigation | parameter A of eq.2 | parameter B of eq.2 |
|---|---|---|
| PrF$_3$ nanoparticles completely covered by $^3$He adsorbed layer (10 cm$^3$ STP). Equilibrium pressure 0.1 mbar. | 8999 | 0.4 |
| PrF$_3$ nanoparticles completely covered by $^3$He adsorbed layer (10 cm$^3$ STP) + gas phase (10 cm$^3$ STP). Equilibrium pressure 25 mbar. | 4140 | 0.4 |
| PrF$_3$ nanoparticles completely covered by $^3$He adsorbed layer (10 cm$^3$ STP) + gas phase (20 cm$^3$ STP). Equilibrium pressure 50 mbar. | 2650 | 0.33 |
| PrF$_3$ nanoparticles + liquid $^3$He. Saturation vapor pressure of $^3$He 75 mbar. | 750 | 0.14 |

Table 1. Parameters A and B for fitting of experimental data (Fig.8) by equation 2.

**M. S. Tagirov[a], E. M. Alakshin[a], R.R. Gazizulin[a], A.V. Egorov[a], A. V. Klochkov[a], S. L. Korableva[a], V.V.Kuzmin[a], A.S.Nizamutdinov[a], K. Kono[b], A. Nakao[b], and A.T. Gubaidullin[c]**

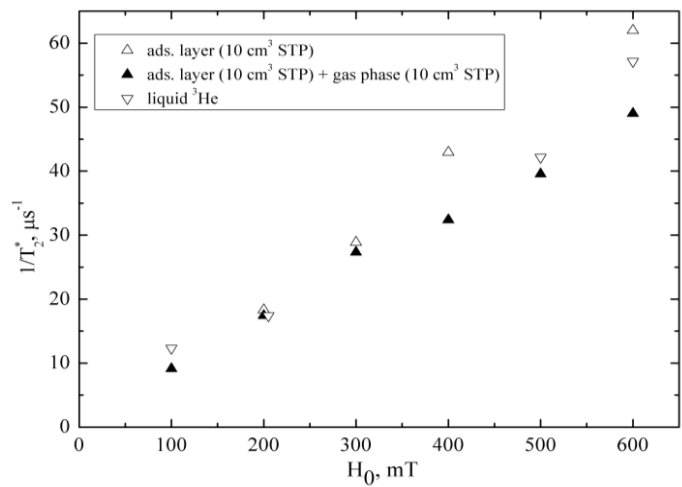

Fig. 9. The dependence of the magnetic field inhomogeneity of the applied external field $H_0$ between $PrF_3$ particles, sensed by $^3He$ $T_2^*$ NMR parameter.

According to fitting experimental data by eq. 2 and parameters presented in table 1 the surface mechanism is quite similar to reported earlier (see for instance [15,17,21]). Moreover, close look at parameter A and its correlation to the amount of gaseous $^3He$ proves the model Hammel-Richardson [23,15] works perfectly in our case and taking into account that intrinsic relaxation time of bulk liquid $^3He$ is much longer than observed values, the model described analytically [18] also works, in spite of significant difference in temperature. Parameter B in the table 1 should be proportional to the diffusion rate of $^3He$, but surprisingly this parameter is the same in the case of complete adsorbed layer of $^3He$ and in the presence of small portion of gaseous (10 cm$^3$ STP) $^3He$, which could be also caused by fast exchange between adsorbed layer and gaseous phase.

## 4. CONCLUSION

The method of synthesis of nanosized powders of crystalline trifluoride rare earths compounds was tested. As a result, two nanoscopic samples of van Vleck paramagnet $PrF_3$ with size $(20 \pm 15)$ nm and $(32 \pm 10)$ nm were synthesized. X-ray analysis established a high crystallinity of the synthesized samples. According to the results of HRTEM, the transition from the first sample to the second sample leads to transition from polycrystalline to single crystal structure.

**NMR of $^3$He in contact with PrF$_3$ nanoparticles at low temperatures**

NMR spectra of $^{141}$Pr in the synthesized PrF$_3$ powders were investigated. The spectrum of nanosized sample is wider than that of microsized PrF$_3$ sample, investigated earlier [12,13]. The simulations of $^{141}$Pr NMR spectra are in good agreement with experimental data.

Spin kinetics of $^3$He in the system "PrF$_3$-$^3$He" was investigated. The model of longitudinal magnetization relaxation of $^3$He nuclei was proposed. According to this model the longitudinal relaxation of $^3$He is carried out both by the $^3$He adsorbed film on the surface and due to the modulation of dipole-dipole interaction in strongly inhomogeneous magnetic field, caused by nanosized PrF$_3$ particles.

At present time according to our experimental data we see no evidence of $^{141}$Pr - $^3$He coupling in our system and it will be the aim of future research.

This work is partially supported by the Ministry of Education and Science of the Russian Federation (FTP "Scientific and scientific - pedagogical personnel of the innovative Russia" GK- P900).